\input lanlmac
\input epsf.tex
\input mssymb.tex
\font\sc=cmcsc10
\input mixedbc.defs
\overfullrule=0pt
\def\figbox#1#2{\epsfxsize=#1%
\vcenter{\hbox{\epsfbox{#2}}}}
\def\raisedfigbox#1#2#3{\epsfxsize=#1%
\raise#3\hbox{\epsfbox{#2}}}
\newcount\figno
\figno=1
\def\fig#1#2#3{%
\xdef#1{\the\figno}%
\writedef{#1\leftbracket \the\figno}%
\nobreak%
\par\begingroup\parindent=0pt\leftskip=1cm\rightskip=1cm\parindent=0pt%
\baselineskip=11pt%
\midinsert%
\centerline{#3}%
\vskip 12pt%
{\bf Fig.\ \the\figno:} #2\par%
\endinsert\endgroup\par%
\goodbreak%
\global\advance\figno by1%
}

\def\e#1{{\rm e}^{#1}}
\def\pre#1{{\tt
arXiv:#1}}
\def\d{{\rm d}}

\def\Rc{{\check R}}

\def\der{\partial}

\newcount\propno
\propno=1
\def\prop#1#2\par{\xdef#1{\the\propno}%
\medbreak\noindent{\sc Proposition \the\propno.\enspace}{\sl #2}\medbreak%
\global\advance\propno by1}
\def\thmn#1#2\par{
\medbreak\noindent{\sc Theorem #1.\enspace}{\sl #2\par}\medbreak%
}
\newcount\thmno
\thmno=1
\def\thm#1#2\par{\xdef#1{\the\thmno}%
\thmn{\the\thmno}{#2}%
\global\advance\thmno by1}
\newcount\conjno
\conjno=1
\def\conj#1#2\par{\xdef#1{\the\conjno}%
\medbreak\noindent{\sc Conjecture \the\conjno.\enspace}{\sl #2}\medbreak%
\global\advance\conjno by1}
\newcount\lemno
\lemno=1
\def\lemma#1#2\par{\xdef#1{\the\lemno}%
\medbreak\noindent{\sc Lemma \the\lemno.\enspace}{\sl #2}\medbreak%
\global\advance\lemno by1}
\def\corol#1\par{%
\medbreak\noindent{\sc Corollary.\enspace}{\sl #1}\medbreak}
\long\def\example#1\par{%
\medbreak\noindent{\sc Example:\enspace}#1\medbreak}
\def\qed{\nobreak\hfill\vbox{\hrule height.4pt%
\hbox{\vrule width.4pt height3pt \kern3pt\vrule width.4pt}\hrule height.4pt}\medskip\goodbreak}
\lref\RS{A.V. Razumov and Yu.G. Stroganov, 
{\sl Combinatorial nature of ground state vector of $O(1)$ loop model},
{\it Theor. Math. Phys.} 
{\bf 138} (2004) 333--337; {\it Teor. Mat. Fiz.} 138 (2004) 395--400, \pre{math.CO/0104216}.}
\lref\BdGN{M.T. Batchelor, J. de Gier and B. Nienhuis,
{\sl The quantum symmetric XXZ chain at $\Delta=-1/2$, alternating sign matrices and 
plane partitions},
{\it J. Phys.} A34 (2001) L265--L270,
\pre{cond-mat/0101385}.}
\lref\dG{J. de Gier, {\sl Loops, matchings and alternating-sign matrices},
{\it Discr. Math.} 298 (2005), 365--388,
\pre{math.CO/0211285}.}
\lref\Rob{D.P.~Robbins, {\sl Symmetry classes of alternating-sign matrices},
\pre{math.CO/0008045}.}
\lref\PRdG{P. A. Pearce, V. Rittenberg and J. de Gier, 
{\sl Critical Q=1 Potts Model and Temperley--Lieb Stochastic Processes},
\pre{cond-mat/0108051}.}
\lref\RSb{A.V. Razumov and Yu.G. Stroganov, 
{\sl $O(1)$ loop model with different boundary conditions and symmetry classes of alternating-sign matrices},
{\it Theor. Math. Phys.} 
{\bf 142} (2005) 237--243; {\it Teor. Mat. Fiz.} 142 (2005) 284--292,
\pre{cond-mat/0108103}.}
\lref\PRdGN{P. A. Pearce, V. Rittenberg, J. de Gier and B. Nienhuis,
{\sl Temperley--Lieb Stochastic Processes}, 
{\it J. Phys. A} {\bf 35 } (2002) L661-L668, \pre{math-ph/0209017}.}
\lref\dGR{J.~de~Gier and V.~Rittenberg,
{\sl Refined Razumov--Stroganov conjectures for open boundaries},
{\it JSTAT} (2004) P09009,
\pre{math-ph/0408042}.}
\lref\MNosc{S. Mitra and B. Nienhuis, {\sl 
Osculating random walks on cylinders}, in
{\it Discrete random walks}, 
DRW'03, C. Banderier and
C. Krattenthaler edrs, Discrete Mathematics and Computer Science
Proceedings AC (2003) 259-264, \pre{math-ph/0312036}.} 
\lref\Kup{G. Kuperberg, {\sl Symmetry classes of alternating-sign matrices under one roof},
{\it Ann. of Math.} (2) 156 (2002), no. 3, 835--866,
\pre{math.CO/0008184}.}
\lref\MNdGB{S. Mitra, B. Nienhuis, J. de Gier and M.T. Batchelor,
{\sl Exact expressions for correlations in the ground state 
of the dense $O(1)$ loop model}, 
{\it JSTAT} (2004) P09010,
\pre{cond-mat/0401245}.}
\lref\DF{P.~Di~Francesco, {\sl 
 A refined Razumov--Stroganov conjecture} I: 
     J. Stat. Mech. P08009 (2004), \pre{cond-mat/0407477}; II: 
     J. Stat. Mech. P11004 (2004), \pre{cond-mat/0409576}.}
\lref\DFb{P.~Di~Francesco,
{\sl Totally Symmetric Self-Complementary Plane Partitions and Quantum Knizhnik-Zamolodchikov equation: a conjecture},
\pre{cond-mat/0607499}.}
\lref\KP{M.~Kasatani and V.~Pasquier,
{\sl On polynomials interpolating between the stationary state of a O(n) model and a Q.H.E. ground state},
\pre{cond-mat/0608160}.}
\lref\LGV{B. Lindstr\"om, {\it On the vector representations of
induced matroids}, Bull. London Math. Soc. 5 (1973)
85--90\semi
I. M. Gessel and X. Viennot, {\it Binomial determinants, paths and
hook formulae}, {\it Adv. Math.} 58 (1985) 300--321. }
\lref\DFZJ{P.~Di Francesco and P.~Zinn-Justin, {\sl Around the Razumov--Stroganov conjecture:
proof of a multi-parameter sum rule}, {\it E. J. Combi.} 12 (1) (2005), R6,
\pre{math-ph/0410061}.}
\lref\Pas{V.~Pasquier, {\sl Quantum incompressibility and Razumov Stroganov type conjectures},
\pre{cond-mat/0506075}, to appear in {\it Ann. Henri Poincar\'e}.}
\lref\FR{I.B.~Frenkel and N.~Reshetikhin, {\sl Quantum affine Algebras and Holonomic Difference Equations},
{\it Commun. Math. Phys.} 146 (1992), 1--60.}
\lref\DFZJb{P. Di Francesco and P. Zinn-Justin,
{\sl Inhomogeneous model of crossing loops and multidegrees of some algebraic varieties},
{\it Commun. Math. Phys.} 262 (2006), 459--487,
\pre{math-ph/0412031}.}
\lref\KZJ{A. Knutson and P. Zinn-Justin,
{\sl A scheme related to Brauer loops},
to appear in {\it Advances In Mathematics},
\pre{math.AG/0503224}.}
\lref\DFZJc{P.~Di Francesco and P.~Zinn-Justin, 
{\sl Quantum Knizhnik--Zamolodchikov equation, generalized Razumov--Stroganov sum rules 
and extended Joseph polynomials}, 
{\it J. Phys. A} 38 (2005) L815--L822, \pre{math-ph/0508059}.}
\lref\DFZJd{P.~Di Francesco and P.~Zinn-Justin, {\sl From Orbital Varieties to Alternating 
Sign Matrices}, to appear in the proceedings of FPSAC'06 (2006), \pre{math-ph/0512047}.}
\lref\DFZJZ{P.~Di~Francesco, P.~Zinn-Justin and J.-B.~Zuber,
{\sl A Bijection between classes of Fully Packed Loops and Plane Partitions},
{\it E. J. Combi.} 11(1) (2004), R64,
\pre{math.CO/0311220}.}
\lref\DFZ{P.~Di~Francesco and J.-B.~Zuber, {\sl On FPL 
configurations with four sets of nested arches}, 
{\it JSTAT} (2004) P06005, \pre{cond-mat/0403268}.}
\lref\DFZJZb{P.~Di~Francesco, P.~Zinn-Justin and J.-B.~Zuber,
{\sl Determinant Formulae for some Tiling Problems and Application to Fully Packed Loops}, 
{\it Annales de l'Institut Fourier} 55 (6) (2005), 2025--2050, \pre{math-ph/0410002}.} 
\lref\Oka{S. Okada, 
{\sl  Enumeration of Symmetry Classes of Alternating Sign Matrices and Characters of Classical Groups}, 
{\it J. Algebr. Comb.} 23 (2006), 43--69,
\pre{math.CO/0408234}.}
\lref\KOR{V. Korepin, {\sl Calculation of norms of Bethe wave functions},
{\it Comm. Math. Phys.} 86 (1982), 391--418.}
\lref\Kratt{F. Caselli and C.~Krattenthaler, {\sl Proof of two conjectures of
Zuber on fully packed loop configurations}, {\it J. Combin. Theory
Ser.} A 108 (2004), 123--146, \pre{math.CO/0312217}.}
\lref\DF{P.~Di Francesco, {\sl Inhomogeneous loop models with open boundaries}, 
{\it J. Phys.} A 38 (2005), 6091, \pre{math-ph/0504032}\semi
{\sl Boundary $q$KZ equation and generalized Razumov--Stroganov sum rules for open IRF models}, 
{\it J. Stat. Mech.} P11003 (2005), \pre{math-ph/0509011}.}
\lref\CZJ{L.~Cantini and P.~Zinn-Justin, work in progres.}
\lref\Lev{D.~Levy, {\sl Algebraic structure of translation-invariant spin-1/2 xxz and q-Potts quantum chains},
{\it Phys. Rev. Lett.} 67 (1991), 1971--1974; {\it Int. J. Mod. Phys.} A6 (1991), 5127--5154.}
\lref\MS{P.~Martin and H.~Saleur, {\sl The Blob Algebra and the Periodic Temperley-Lieb Algebra},
{\it Lett. Math. Phys.} 30 (1994), 189, \pre{hep-th/9302094}.}
\lref\Skl{E. K.~Sklyanin, {\sl Boundary conditions for integrable quantum systems},
{\it J. Phys.} A 21 (1988), 2375--2389.}
\lref\IO{A.~Isaev and O.~Ogievietsky, {\sl Baxterized Solutions of Reflection Equation and Integrable Chain Models},
{\it Nucl. Phys.} B 760 [PM] (2007), 167--183, \pre{math-ph/0510078}.}
\Title{}
{
\vbox{\centerline{Loop model with mixed boundary conditions,}
\medskip
\centerline{$q$KZ equation and Alternating Sign Matrices}}
}
\bigskip\bigskip
\centerline{P. Zinn-Justin \footnote{${}^\star$}
{Laboratoire de Physique Th\'eorique et Mod\`eles Statistiques
(CNRS, UMR 8626); Univ Paris-Sud, Orsay, F-91405.}}
\vskip0.5cm
\noindent
The integrable loop model with mixed boundary conditions based on the 1-boundary extended Temperley--Lieb
algebra with loop weight 1 is considered. The corresponding $q$KZ equation is introduced and its minimal degree solution
described. As a result, the sum of the properly normalized components of the ground state in size $L$ is computed 
and shown to be equal to the number of Horizontally and Vertically Symmetric Alternating Sign Matrices
of size $2L+3$. A refined counting is also considered.

\bigskip

\def\Rc{{\check R}}

\Date{10/2006}
%
%
%
\newsec{Introduction}
The present work is yet another chapter in the continuing story,
inspired by the seminal papers \refs{\BdGN,\RS},
of the interrelation between quantum integrability and combinatorics.
It concerns the combinatorial interpretation of the ground state components of a quantum integrable 1D Hamiltonian,
or equivalently of the equilibrium distribution of a Markov process on planar connectivities of points (also called
link patterns). We refer the reader to \dG\ for an overview of the relevant model, related to the Temperley--Lieb alegbra.
These ground state entries turn out to be integers which count classes of Fully Packed Loop (FPL) configurations, 
better known in the mathematical literature as Alternating Sign Matrices (ASMs).
An important detail is that while the first articles \refs{\BdGN,\RS} considered models with periodic boundary conditions,
subsequent work also studied also types of boundaries conditions \refs{\PRdG,\RSb,\PRdGN,\MNdGB}. In particular
in \refs{\dG,\MNdGB} one can find a definition of the so-called mixed boundary conditions which will be used here.
In the loop model to ASM dictionary, boundary conditions of the model translate into symmetry classes of ASMs.

Here we make use of the method of study developed in \DFZJ, and of its extension considered in \Pas, or more precisely
of its reformulation in terms of $q$KZ equation given in \DFZJc. The essence of the method is to generalize
the problem by introducing extra parameters (in a similar spirit as \Kup), making the model
inhomogeneous. We go further by replacing the eigenvector equation with a more general linear system, 
which we call a bit loosely the $q$KZ equation,
which lets us vary an extra parameter $q$. The latter generalization, although not strictly necessary for our purposes,
is both conceptually satisfactory (allowing to distinguish clearly what properties are dependent or independent of $q$),
and of some mathematical interest, in relation to representation theory \Pas\ and algebraic geometry \DFZJc.

The plan of the article is as follows: in Sect.~2 we give a brief description of the loop model with mixed boundary
conditions and define our notations. In Sect.~3 we define the $q$KZ equation and discuss the relevant solution.
Sect.~4 contains the main application of this formalism: the computation of the sum rule of the model with
mixed boundary conditions, as well as its refinement in the sense of \dGR. Sect.~5 provides concluding comments.

\newsec{Definition of the model}
We provide here the minimum about the model with mixed boundary conditions
introduced in \dG\ that is needed for our purposes, as well as other useful definitions.
Let $L$ be a fixed positive integer. Consider points on a line,
numbered from $1$ to $L$, plus
an extra point to the right of them.
The space of states consists of {\it right-extended link patterns}\/ (or
right-extended matchings, in the language of \dG), that is pairings of
the points $\{1,\ldots,L\}$ between themselves, allowing
for an arbitrary number of unpaired points which are connected
to the extra rightmost point, in such a way that all these
connections can be drawn in the half-plane without crossings, see
Fig.~\exlp. There are ${L\choose\lfloor L/2 \rfloor}$ such link patterns.

\fig\exlp{The right-extended link patterns of size $L=3,4$.}{%
\vbox{\hbox{$L=3$:}\vskip1.4cm\hbox{$L=4$:}\vskip1.7cm}
\vbox{
\hbox{
\epsfxsize=4cm\epsfbox{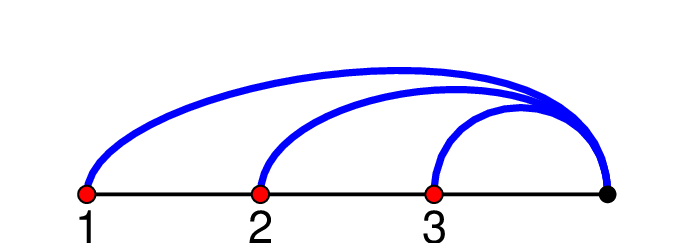}
\epsfxsize=4cm\epsfbox{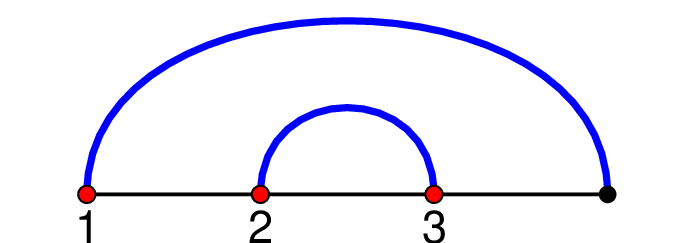}
\epsfxsize=4cm\epsfbox{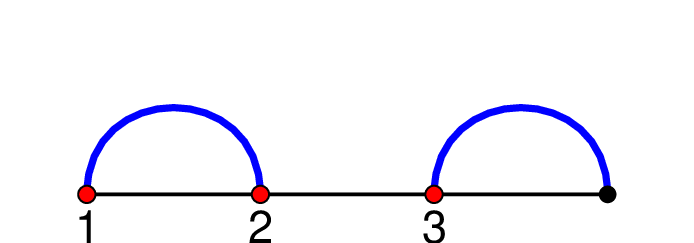}
}
\vskip10pt
\hbox{
\epsfxsize=4cm\epsfbox{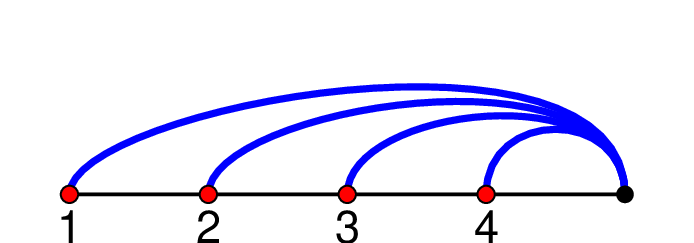}
\epsfxsize=4cm\epsfbox{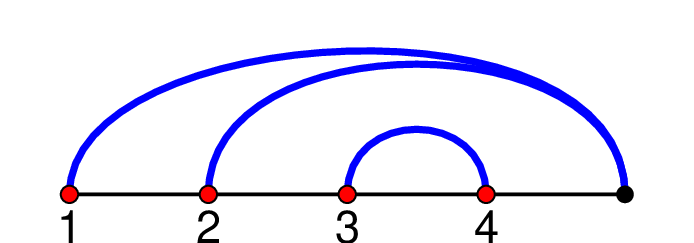}
\epsfxsize=4cm\epsfbox{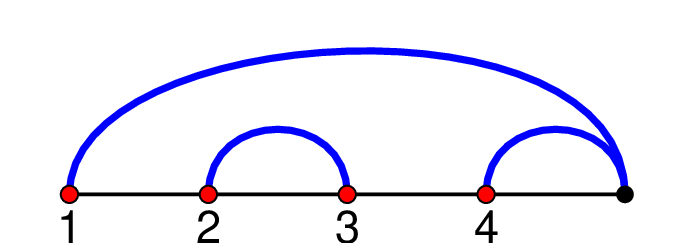}
}
\hbox{
\epsfxsize=4cm\epsfbox{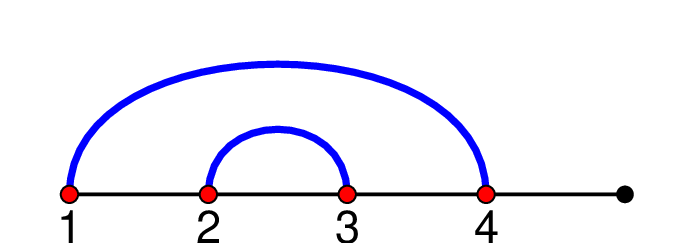}
\epsfxsize=4cm\epsfbox{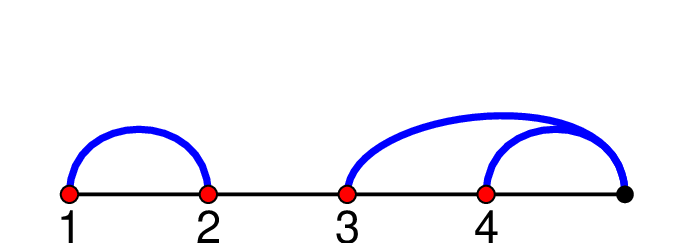}
\epsfxsize=4cm\epsfbox{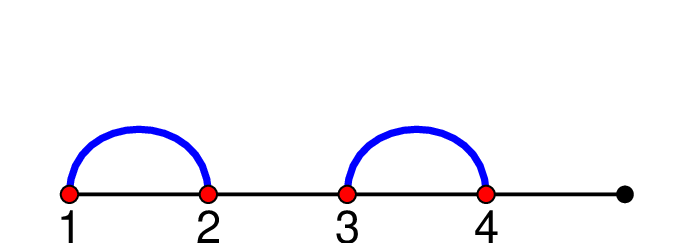}
}}
}

These link patterns form the canonical basis of a vector space, on
which the one-boundary-extended
Temperley--Lieb algebra \refs{\Lev,\MS} acts. Its generators,
$e_1,\ldots e_{L-1}$ and $f$ send link patterns to other link patterns
according to the standard graphical rule 
exemplified on Fig.~\extl. Each time a loop is formed, it must
be erased and the state must be multiplied by a weight
$\tau$. On the other hand ``boundary loops'' that go through
the rightmost point produce no weight.
The operators $e_i$ and $f$ satisfy the following relations
\eqnn\tlrel
$$\eqalignno{
&e_i^2=\tau e_i\qquad e_ie_{i\pm 1}e_i=e_i\qquad e_ie_j=e_je_i\quad |i-j|>1\cr
&f^2=f\qquad e_{L-1}f e_{L-1}=e_{L-1}\qquad e_i f=f e_i\quad i<L-1
&\tlrel\cr
}$$
\fig\extl{Examples of action of Temperley--Lieb operators.}{%
\vbox{
\hbox{$e_1 \raisedfigbox{4cm}{diag4-3.eps}{-0.2cm} 
= \raisedfigbox{4cm}{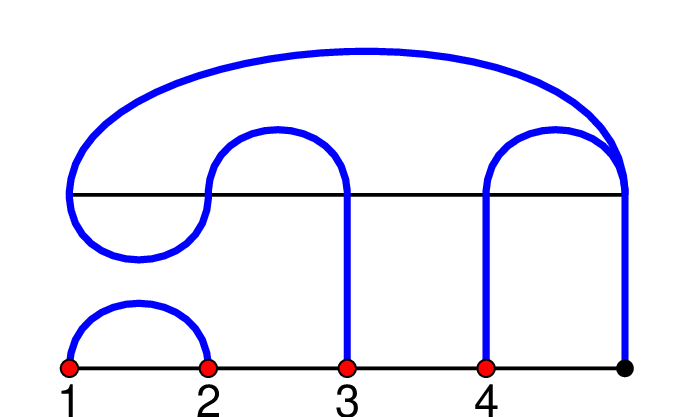}{-0.5cm} 
= \raisedfigbox{4cm}{diag4-5.eps}{-0.2cm}$}
\hbox{$e_2 \raisedfigbox{4cm}{diag4-6.eps}{-0.2cm} 
= \raisedfigbox{4cm}{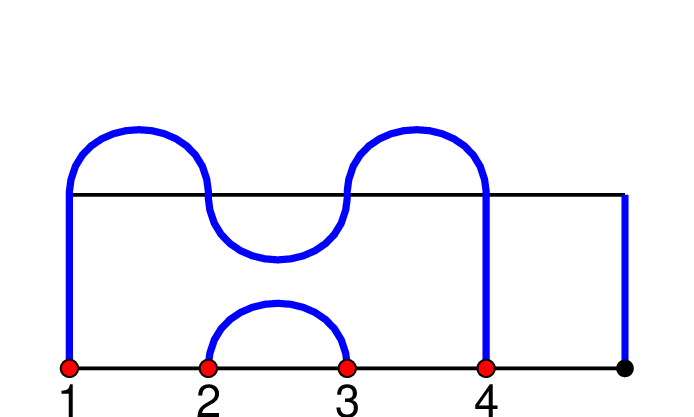}{-0.5cm} 
= \raisedfigbox{4cm}{diag4-4.eps}{-0.2cm}$}
\hbox{$e_3 \raisedfigbox{4cm}{diag4-2.eps}{-0.2cm} 
= \raisedfigbox{4cm}{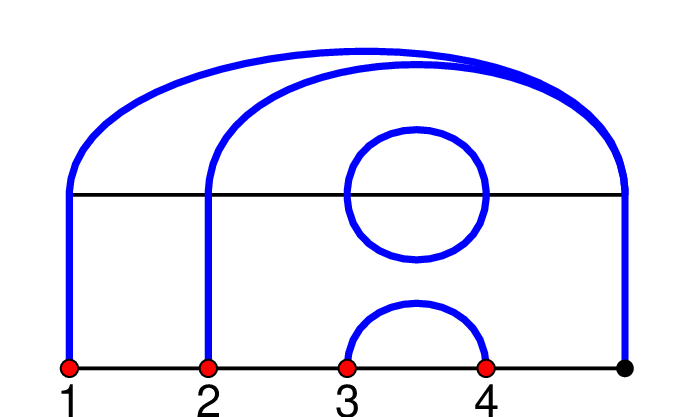}{-0.5cm} 
=\ \tau \raisedfigbox{4cm}{diag4-2.eps}{-0.2cm}$}
\hbox{$f_{\ } \raisedfigbox{4cm}{diag4-6.eps}{-0.2cm} 
= \raisedfigbox{4cm}{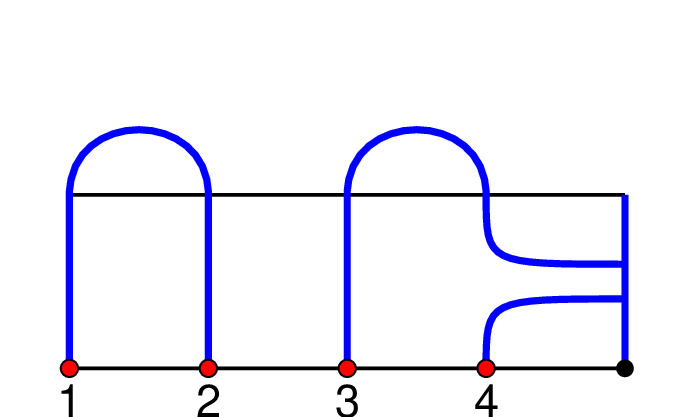}{-0.5cm} 
= \raisedfigbox{4cm}{diag4-5.eps}{-0.2cm}$}
}
}

The term ``mixed boundary conditions'' refers to the fact that
link patterns are closed on the left, but open/extended on the right.

Set $\tau=-q-q^{-1}$. One can define the $R$-matrix
\eqn\Rmat{
\Rc_i(z)={(q\,z-1/q)I+(z-1)e_i\over q-z/q}
}
where $I$ is the identity operator,
and the boundary $R$-matrix \refs{\Skl,\IO}
\eqn\Rbmat{
K(z)={(z-q^2/\zeta)(z-\zeta/q)I
+(1-q)(1-z^2)f\over (q\, z-\zeta/q)(z-q/\zeta)}
}
where $\zeta$ is an additional free parameter of the model (this expression can also be found in a
different parameterization in the appendix of \MNdGB).

These operators satisfy the unitarity equation:
\eqn\unit{
\Rc_i(z)\Rc_i(1/z)=I\qquad 
K(z)K(1/z)=I
}
the Yang--Baxter equation:
\eqn\YBE{
\Rc_i(z)\Rc_{i+1}(z\, w)\Rc_i(w)=\Rc_{i+1}(w)\Rc_i(z\, w)\Rc_{i+1}(z)
}
and the boundary Yang--Baxter (or reflection) equation: 
\eqn\BYBE
{
\Rc_{L-1}(w/z)
K(z)
\Rc_{L-1}(1/(wz))
K(w)
=
K(w)
\Rc_{L-1}(1/(wz))
K(z)
\Rc_{L-1}(w/z)
}
as well as the usual commutation relations $[\Rc_i(z),\Rc_j(z')]=0$
for $|i-j|>1$ and $[K(z),\Rc_i(z')]=0$ for $i<L-1$.

An integrable model is usually defined by a infinite family
of commuting transfer matrices. In the present context it is more relevant to
consider {\it scattering matrices}\/ $S_i$, $i=1,\ldots
L-1$, defined by
\eqnn\defS
$$\eqalignno{
S_i(z_1,\ldots,z_L)=
&\Rc_i(s\,z_i/z_{i+1}) \Rc_{i+1}(s\,z_i/z_{i+2})\ldots \Rc_{L-1}(s\,z_i/z_L)\cr
&K(1/(s\,z_i))\cr
&\Rc_{L-1}(s\,z_L z_i)\ldots\Rc_i(s\,z_{i+1}z_i) \Rc_{i-1}(s\,z_{i-1}z_i)\ldots  \Rc_1(s\,z_1 z_i)\cr
&\Rc_1(z_i/z_1)\ldots \Rc_{i-1}(z_i/z_{i-1})
&\defS}
$$
where $s$ is yet another free parameter.
Intuitively (as in the Coordinate Bethe Ansatz),
$S_i$ corresponds to a ``particle'' starting at location $i$ with spectral parameter $z_i$,
moving to the left and crossing other particles with spectral parameters $z_j$, $j<i$, then reflecting at the left
boundary by having its spectral parameter changed to $1/(s\,z_i)$ (but with a trivial left boundary $R$-matrix),
then going back to the right crossing all other particles, reflecting to the right with the matrix $K$ and having
its spectral parameter inverted to $s\,z_i$, and finally moving back left to its original location by crossing
particles with spectral parameters $z_j$, $j>i$.

It is a simple exercise to show, using the properties above, that
these operators $S_i$ satisfy the following commutation relations:
\eqn\commS{
S_i(z_1,\ldots,s\, z_j,\ldots,z_L)S_j(z_1,\ldots,z_L)
=
S_j(z_1,\ldots,s\, z_i,\ldots,z_L)S_i(z_1,\ldots,z_L)
}
In particular note that when $s=1$, the $S_i$ commute. This is the ``physical'' situation that
leads to the diagonalization problem for the $S_i$ (which will be investigated in detail in Sec.~4).
However, for the time being, we do want to keep $s$ free.

\newsec{$q$KZ equation and its Laurent polynomial solution}
\subsec{The linear system}
\eqna\qkz
Consider the following set
of equations for $\Psi_L$, a Laurent polynomial in
the variables $z_1,\ldots,z_L$:\foot{$\Psi_L$ can be made into a polynomial
by multiplying it by $(z_1\cdots z_L)^k$ for $k$ sufficiently large; this however
makes Eqs.~\qkz{b,c} 
look slightly less nice.} with values in the vector space spanned by the right-extended link patterns of size $L$:
$$\eqalignno{
\Rc_i(z_{i+1}/z_i)\Psi_L(z_1,\ldots,z_i,z_{i+1},\ldots,z_L)
&=
\Psi_L(z_1,\ldots,z_{i+1},z_i,\ldots,z_L)
&\qkz{a}\cr
K(z_L)
\Psi_L(z_1,\ldots,z_L)
&=
\Psi_L(z_1,\ldots,z_{L-1},1/z_L)
&\qkz{b}\cr
\Psi_L(z_1,\ldots,z_L)
&=
\Psi_L(1/(s\,z_1),z_2,\ldots,z_L)
&\qkz{c}\cr
}$$

These equations mimic the definition of the model, corresponding to the behavior of the
ground state eigenvector $\Psi_L$ at the bulk, the right boundary and the left boundary respectively.
In order to explain more precisely the connection with the previous section, one
can compute $S_i \Psi_L$ for $\Psi_L$ a solution of equations \qkz{}.
Going back to the definition \defS\ of the operators $S_i$,
we apply successively Eqs.~\qkz{a} for $i,i+1,\ldots,L-1$, then Eq.~\qkz{b},
then Eqs.~\qkz{a} for $L-1,\ldots,1$, then Eq.~\qkz{c}, then Eqs.~\qkz{a} for $1,\ldots,i-1$
and the end result is
\eqn\realqkz{
S_i(z_1,\ldots,z_L)\Psi_L(z_1,\ldots,z_L)=
\Psi_L(z_1,\ldots,s\, z_i,\ldots,z_L)
}
Thus, $\Psi_L$ is a vector that has very simple transformation properties under the action of
the $S_i$. Eq.~\realqkz\ is in fact the $q$KZ equation
(or more precisely, a variant of it since 
the original $q$KZ equation \FR\ is really the analogue for periodic boundary equations).

\subsec{Laurent polynomial solution}
We now claim the following:

{\sl If $s=q^3$, there exists a solution of
Eqs.~\qkz{}, which is a centered Laurent polynomial of total degree width $2 L(L-1)/2$ and degree width in each
variable $2(L-1)$. The solution is unique up to normalization and
can be constructed explicitly by starting from the base component
(empty link pattern with no pairings, first in the examples of Fig.~\exlp)
\eqn\basepsi{
\Psi_{L;0}=\prod_{1\le i<j\le L} (q\,z_i z_j^{-1} - q^{-1})(q^{-1}z_j-qs^{-1}z_i^{-1})
}
and applying Eqs.~\qkz{a,b}.}

Similar statements have been made in various related models, and the
strategy to prove them always follows the same general pattern. 
Here we shall simply explain why the equality $s=q^3$, as well as Eq.~\basepsi, necessarily hold, 
and how the latter fixes all other components.

First we rewrite Eqs.~\qkz{a,b} by using the explicit forms (Eqs.~\Rmat, \Rbmat) of $\Rc$ and $K$.
We obtain the following dichotomies:

\item{a1)} If the link pattern $\pi$ contains the pairing $(i,i+1)$, then noting that $e_i\pi=\tau\pi$
we can rewrite the component $\pi$ of Eq.~\qkz{a} as
\eqn\dda{
\sum_{\pi'\ne\pi: e_i\pi'=\pi} \Psi_{\pi'}=(q\,z_i-q^{-1}z_{i+1})\, \der_i \Psi_\pi
}
where $\der_i$ is the divided difference operator: 
$\der_i F(z_i,z_{i+1})={\textstyle F(z_{i+1},z_i)-F(z_i,z_{i+1})\over\textstyle z_{i+1}-z_i}$.

\item{a2)} If $i$ and $i+1$ are not paired in $\pi$, then $\Psi_\pi|_{z_{i+1}=q^2 z_i}=0$ and
$\Psi_\pi/(q\,z_i - q^{-1}z_{i+1})$ is symmetric in $z_i$, $z_{i+1}$.

and:

\item{b1)} If the point $L$ is unpaired in $\pi$, then noting that $f\pi=\pi$ 
we can rewrite the component $\pi$ of Eq.~\qkz{b} as
\eqn\ddb{
\sum_{\pi'\ne\pi: f\pi'=\pi} \Psi_{\pi'}={1\over 1-q} 
(q-q^{-1}\zeta z_L^{-1})(z_L-q\,\zeta^{-1})
\, \tilde\der \Psi_\pi
}
where $\tilde\der F(z_L)={\textstyle F(1/z_L)-F(z_L)\over\textstyle 1/z_L-z_L}$.

\item{b2)} If the point $L$ is paired in $\pi$ (i.e.\ not connected to
the rightmost point), then $\Psi_\pi$ has zeroes at $z_L=q^{-2}\zeta$,
$z_L=q\,\zeta^{-1}$. 
and $\Psi_\pi/((q-q^{-1}\zeta z_L^{-1})(z_L-q\,\zeta^{-1}))$ 
is unchanged by $z_L\to 1/z_L$.

Using recursively these properties, as well as the remaining equation \qkz{c},
one can determine all the ``trivial'' factors in the components
$\Psi_\pi$ depending on the properties of $\pi$:

\noindent $\star$ If no pairings occur between the consecutive points $\{i,\ldots,j\}$,
$\prod_{i\le k<l\le j} (q\,z_k-q^{-1}z_l)\, |\, \Psi_\pi$.

\noindent $\star$ If no pairings occur between the consecutive points $\{1,\ldots,j\}$,
$\prod_{1\le k<l\le j} 
(q^2-s\, z_k z_l)\, |\, \Psi_\pi$.

\noindent $\star$ If $L$ is unpaired and there are no pairings between the consecutive points
$\{i,\ldots,L-1\}$ and no unpaired points among them, 
$\prod_{i\le k<l\le L} 
(1-q^2z_k z_l)\,|\,\Psi_\pi$.

\noindent $\star$ If $L$ is paired and there are no pairings in the interval $\{i,\ldots,L\}$,
$\prod_{k=i}^L
(q-q^{-1}\zeta z_k^{-1})(z_k-q\,\zeta^{-1})
\,|\,\Psi_\pi$.

The first and second properties imply that $\Psi_{L;0}$ contains all the factors of Eq.~\basepsi, and since
these exhaust the degree width of the claim, there can be no more (monomials included, since $\Psi_{\pi_0}$
must be centered). 
Furthermore, it is clear that Eqs.~\dda\
and \ddb\ allow to build all other components starting from $\Psi_{L;0}$, and can only preserve or 
lower the degree width.
Finally, we shall show explicitly in the following examples how $s=q^3$ is necessary for $L=2$, $L=3$. 
Below we shall exhibit recurrence relations
that allow to produce $\Psi_{L-2}$ from $\Psi_L$; so that the condition $s=q^3$ must be true for any $L$.

\example{In size $L=2$, there are only two components: the empty link pattern (numbered $0$) and
the link pattern that pairs $1$ and $2$ (numbered $1$). Starting from
$$\Psi_{2;0}=(qz_1z_2^{-1}-q^{-1})(q^{-1}z_2-qs^{-1}z_1^{-1})$$
and noting that $f$ sends $1$ to $0$, we compute
$$\Psi_{2;1}={1\over 1-q}
(q-q^{-1}\zeta z_2^{-1})(z_2-q\,\zeta^{-1})
\tilde\der\Psi_{2;0}={(q^4-s)(q\,z_2-\zeta/q)(z_2-q/\zeta)\over(1-q)q^2s\,z_2}$$
But we also have that $e_1$ sends $0$ to $1$ so that we can also compute
$$0=\Psi_{2;0}-(q\,z_1-q^{-1}z_2)\der_1\Psi_{2,1}={(s-q^3)(q^2z_1-z_2)(1-q^2z_1z_2)\over(1-q)q^3s\,z_1z_2}$$
so that we conclude that $s=q^3$. Substituting this leads to
$$\eqalignno{
\Psi_{2;0}&=(z_1z_2^{-1}-q^{-2})(z_2-q^{-1}z_1^{-1})\cr
\Psi_{2;1}&=-q^{-2}(q-q^{-1}\zeta z_2^{-1})(z_2-q\,\zeta^{-1})
\cr
}$$
(note in particular the simplification in $\Psi_{2;1}$ which becomes a Laurent polynomial in $q$).

In size $L=3$, we use the ordering of Fig.~\exlp. Starting from $\Psi_{3;0}$, we compute 
$\Psi_{3;1}={1\over 1-q}(q-q^{-1}\zeta z_3^{-1})(z_3-q\,\zeta^{-1})
\tilde\der\Psi_{3;0}$ and 
$\Psi_{3;2}=-\Psi_{3,0}+(q\,z_2-q^{-1}z_3)\der_2\Psi_{3;1}$; skipping these intermediate steps, we note that since $e_1^{-1}(2)=\{0,1,2\}$, we must have
$$0=\Psi_{3;0}+\Psi_{3;1}-(q\,z_1-q^{-1}z_2)\der_1\Psi_{3,2}=(s-q^3)(q^2z_1-z_2)(q^2z_1-z_3)(q^2z_2-z_3)R(z_1,z_2,z_3)$$
where $R$ is some irrelevant rational fraction. We conclude again that $s=q^3$, and after simplification:
$$\eqalign{
\Psi_{3;0}=&(z_1z_2^{-1}-q^{-2})(z_2-q^{-1}z_1^{-1})(z_1z_3^{-1}-q^{-2})(z_3-q^{-1}z_1^{-1})(z_2z_3^{-1}-q^{-2})(z_3-q^{-1}z_2^{-1})\cr
\Psi_{3;1}=&(z_1z_2^{-1}-q^{-2})(z_2-q^{-1}z_1^{-1})
(q-q^{-1}\zeta z_3^{-1})(z_3-q\,\zeta^{-1})
\times\cr
&\times(-q^{-2}(z_1+z_2)+q^{-3}(1+q^{-1})z_3-q^{-5}(z_1^{-1}+z_2^{-1})+q^{-3}(1+q^{-1})z_3^{-1})\cr
\Psi_{3;2}=&(z_2z_3^{-1}-q^{-2})(z_3-q^{-2}z_2^{-1})
({\scriptstyle -q^{-1}z_1 (z_2+z_3) +q^{-2}(1+q^{-1})z_2 z_3 
+(q^{-3}\zeta+\zeta^{-1})((1+q)z_1-q^{-1}(z_2+z_3))}\cr
&{\scriptstyle -q^{-1}z_1 (z_2^{-1}+z_3^{-1})-q^{-4}(z_2+z_3) z_1^{-1} 
+q^{-2}(1+q^{-1})(z_2 z_3^{-1}+z_3 z_2^{-1})
+(q^{-3}\zeta+\zeta^{-1})(q^{-2}(1+q^{-1})z_1^{-1}-q^{-1}(z_2^{-1}+z_3^{-1}))}\cr
&{\scriptstyle -(1-q^{-1})^2q^{-1}(1+q^{-1})
-q^{-4}z_1^{-1}(z_2^{-1}+z_3^{-1})+q^{-2}(1+q^{-1})z_2z_3^{-1}
})\cr 
}$$}

\subsec{Recurrence relations}
We derive here a recurrence relation with respect to the size $L$ satisfied by $\Psi_L$.

Consider $\Psi_L|_ {z_{i+1}=q^2 z_i}$ for $1\le i\le L-1$. Using the results of previous section we know that
only the components with a pairing $(i,i+1)$ are non-zero. Defining the operation $\phi_i$ that inserts into a link
pattern $\pi$ of size $L-2$ the pairing $(i,i+1)$ (thus shifting the points $i,\ldots,L-2$ by $2$), we claim that
\eqn\reca{
\Psi_{L,\phi_i(\pi)}(z_1,\ldots,z_L)|_{z_{i+1}=q^2 z_i}=P_i(z_i;z_1,\ldots,z_{i-1},z_{i+2},\ldots,z_L)\,
\Psi_{L-2,\pi}(z_1,\ldots,z_{i-1},z_{i+2},\ldots,z_L)
}
where $P_i$ is some Laurent polynomial of degree width $2(2L-3)$ which is symmetric in $z_1,\ldots,z_{i-1},z_{i+2},z_L$.
The common factor $P_i$ can be found by factor exhaustion, up to a multiplicative constant; 
it is then simple to check that the l.h.s.\ of Eq.~\reca\ 
satisfies Eqs.~\qkz{} at size $L-2$; one concludes by uniqueness of the solution of the prescribed degree that
Eq.~\reca\ holds.
Remains a multiplicative constant that needs to be computed by taking an example.

Since we shall need in what follows only one recurrence relation, we do the computation of $P_i$ only for $i=L-1$.
In this case note that the empty link pattern of size $L-2$ is sent by $\phi_{L-1}$ to the link pattern
with one pairing $(L-1,L)$. The latter is the only antecedent of the empty link pattern by $f$; thus
its component is
$$\Psi_{L,1}=
{1\over 1-q}
(q-q^{-1}\zeta z_L^{-1})(z_L-q\,\zeta^{-1})
\tilde\der \Psi_{L,0}$$
Setting $z_L=q^2 z_{L-1}$ we note that $\Psi_{L,0}$ vanishes, so that we are left with only one term
in $\tilde\der\Psi_{L,0}$. Explicitly,
$$\eqalign{
\Psi_{L,1}|_{z_L=q^2 z_{L-1}}=&
(q\,z_{L-1}-\zeta^{-1})(1-q^{-4}\zeta z_{L-1}^{-1})
\prod_{1\le i<j\le L-1} (z_i z_j^{-1} - q^{-2})(z_j-q^{-1}z_i^{-1})
\times\cr 
&\times\prod_{1\le i\le L-2}(z_iz_{L-1}^{-1}-q)(z_{L-1}-q^{-4}z_i^{-1})
}$$
Dividing by $\Psi_{L-2,0}(z_1,\ldots,z_{L-2})$ we obtain
\eqnn\exP
$$\eqalignno{
P_{L-1}(z_{L-1};z_1,\ldots&,z_{L-2})=(q\,z_{L-1}-\zeta^{-1})(1-q^{-4}\zeta z_{L-1}^{-1})
\times&\exP\cr
&\times
\prod_{i=1}^{L-2} (z_i z_{L-1}^{-1} - q^{-2})(z_{L-1}-q^{-1}z_i^{-1})
(z_iz_{L-1}^{-1}-q)(z_{L-1}-q^{-4}z_i^{-1})
}$$
All the factors can be expected on general grounds (zeroes of $\Psi_L$), but our computation has produced the
normalization constant.

\newsec{Sum rule at the stochastic point}
\subsec{Stochastic point}
A value of the parameters that is of special interest is
$q=\e{\pm 2 i\pi/3}$,
that is the weight of a loop is $\tau=1$, and $s=q^3=1$.
At this special value we notice that all operators
$e_i$ and $f$ have the co-vector $v=(1,\ldots,1)$ as left eigenvector,
with eigenvalue $1$:
$v e_i=v f=v$. So do the operators $\Rc_i$, $K$, as well as $S_i$.
In stochastic processes, this is interpreted as the conservation
of probability. Also, we note that the operators $S_i$ now commute according to
Eq.~\commS.
It will therefore come as no surprise that Eq.~\realqkz\ becomes the equation for their common (right) eigenvector with
eigenvalue $1$:
\eqn\defPsiz{
S_i(z_1,\ldots,z_L)\Psi_L(z_1,\ldots,z_L)=\Psi_L(z_1,\ldots,z_L)
}

We can now consider the limit $z_i\to 1$, in which all operators
$S_i$ tend to the identity. Expanding to first order in any variable
we note that ${\der\over\der z_j} S_i|_{z_k=1}=\alpha_{ij}
H+\beta_{ij}$, 
where $\alpha_{ij}$ and $\beta_{ij}$ are irrelevant constants,
and $H$ is the Hamiltonian
\eqn\Ham{
H=\sum_{i=1}^{L-1} e_i +a\,f
}
in which we have set 
$a=3/(\zeta+1+\zeta^{-1})$. 
$H$, being a sum of $L$ operators with the same left eigenvector, also possesses it, with eigenvalue
$L-1+a$.
Furthermore, if $a$ is real positive,
it is easy to show that
the matrix of $H$ in the canonical basis
satisfies all properties of the Perron--Frobenius theorem,
so that $L-1+a$ is its largest eigenvalue,
and its right eigenvector $\Psi_L$ defined by
\eqn\defPsi{
H\Psi_L=(L-1+a)\Psi_L
}
is its Perron--Frobenius eigenvector. Of course $\Psi_L$
is up to a multiplicative constant $\Psi_L(1,\ldots,1)$, but we choose its
normalization in such a way that the component indexed by the
empty link pattern is $a^{\lfloor L/2\rfloor}$.

Let us first discuss the case $\zeta=1$, i.e.\ $a=1$.
All componenents turn out to be integers, and in \dG\ is given a
remarkable conjectural combinatorial interpretation of these numbers
(\`a la Razumov--Stroganov conjecture \RS) which we explain now.

Define a Fully Packed Loop configuration (FPL) to be a coloring of edges
of a square lattice in two colors (say, occupied and empty) in such a way that
external edges are alternatingly occupied and empty and that vertices have exactly two
occupied adjacent edges and two empty adjacent edges. FPLs are known to be
in bijection with Alternating Sign Matrices (ASMs), 
see \dG\ for details. A FPL is called vertically
(resp.\ horizontally) symmetric if it is symmetric with respect to the vertical
(resp.\ horizontal) axis. Now we can state the conjecture:
with our normalization each component $\Psi_\pi$ is an integer
and counts the number of Horizontally and Vertically Symmetric
Fully Packed Loop configurations of size $2L+3$
with the connectivity of the corresponding link
pattern $\pi$, see Fig.~\hvsfpl. As a corollary, the sum of components 
$v\cdot\Psi=\sum_\pi \Psi_\pi$ is
equal to the total number of such configurations, or equivalently
the number of Horizontally and Vertically Symmetric Alternating
Sign Matrices (HVSASMs).

\fig\hvsfpl{Example of a HVSFPL and the corresponding link pattern.}{
$\figbox{3.5cm}{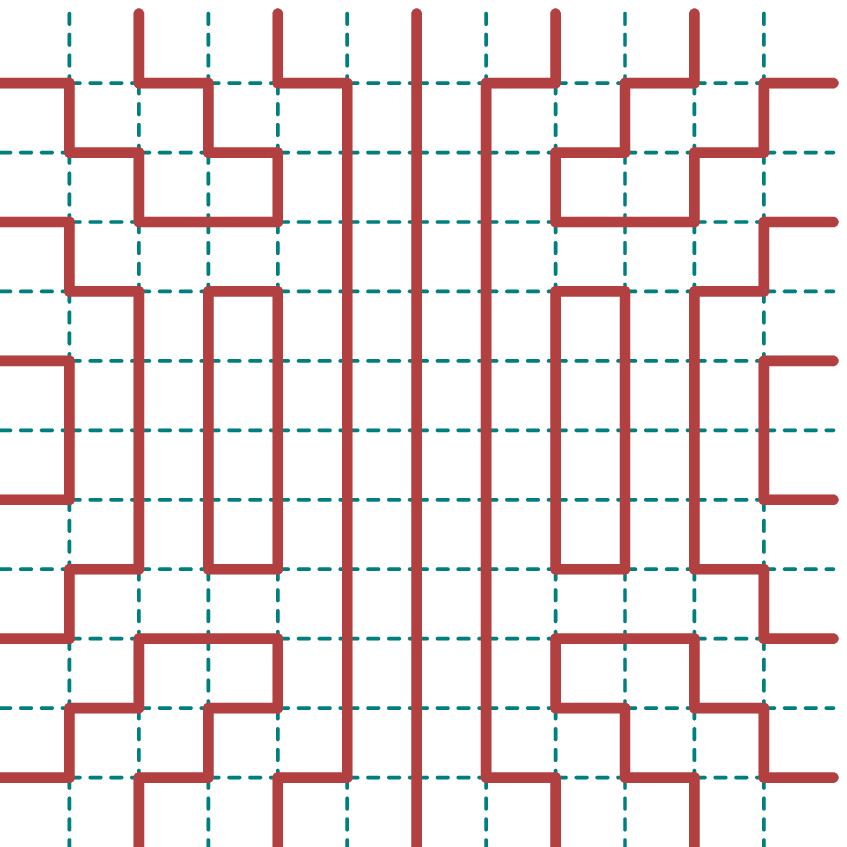}
\qquad\longrightarrow\qquad
\figbox{4cm}{diag4-3.eps}$
}

Let us now go back to a general value of $a$. A refinement of the
previous conjecture, formulated in \dGR, states that the components
provide an $a$-enumeration of the HVSFPLs, the power $a^{\lfloor k/2\rfloor}$ 
being related to the number $k$ of paths crossing the horizontal
symmetry axis say, to the left of the center, that go straight down
two steps below this axis.
Again, the sum of components becomes the $a$-enumeration of HVSASMs, 
where a weight $a$ is given to each pair of $-1$'s 
on the row below the horizontal symmetry axis and to the left of the center.

{\it Remark}: at $a=0$ the Hamiltonian becomes that of the model with closed boundary conditions
(whose ground state eigenvector is related to the counting of VSASMs or even-sized analogues, see \refs{\PRdG,\PRdGN,\dG}),
and one expects to recover the results of \DF, which discusses the inhomogeneous generalization. 
Keeping only the leading term of $\Psi$ as $\zeta\to0$, one finds
that only components with all points paired (except one if $L$ is odd) are non-zero, which is the sector preserved by
the operators $e_i$, $i=1,\ldots,L-1$; and one could naively expect these limiting components 
to be exactly the polynomials introduced in \DF, up to a common factor. 
Interestingly, it is not the case -- if only
because the polynomials in our case have higher degree and are coprime. A related crucial mismatch is that
the $q$KZ equations are different: that of the closed boundary conditions has a shift $s=q^6$, as opposed to $s=q^3$ here.
Only at $q=\e{\pm 2i\pi/3}$, where the two $\Psi$ are ground state eigenvectors of the same transfer matrix
or scattering matrices, is the identification correct. Indeed one can check that the components of
$\Psi_L(\zeta\to0)$  develop a common symmetric factor at $q=\e{\pm 2i\pi/3}$ 
(which is precisely the function $\chi_L(z_1,\ldots,z_L)$ to be
defined below, cf Eq.~\defchi).
A similar phenomenon of interrelation between $s=q^3$ and $s=q^6$ solutions of $q$KZ was observed in \refs{\Pas,\KP} in
the case of periodic boundary conditions.

\example{For $L=4$, with the order of link patterns as on Fig.~\exlp,
\eqn\exH{
H=\pmatrix
{
 a & a & 0 & 0 & 0 & 0 \cr
 1 & 1 & 1 & 0 & 0 & 0 \cr
 1 & 1 & 1+a & a & 1 & 0 \cr
 0 & 0 & 0 & 1 & 0 & 1 \cr
 1 & 0 & 1 & 0 & 1+a & a \cr
 0 & 1 & 0 & 2 & 1 & 2
}
}
The sum of each column is $3+a$, as should be, and $\Psi_4$ is
\eqn\exPsi{
\Psi_4=(a^2,3a,2a(3+a),3,3a(2+a),3(2+a))
}
The corresponding HVSASMs are drawn in \dGR.
At $a=1$, $\Psi_4=(1,3,8,3,9,9)$,
which sums to $33$, the total number of HVSASMs of size $11$.
At $a=0$, $\Psi_4=3(0,0,0,1,0,2)$, which is the ground state eigenvector of the Hamiltonian
with closed boundary conditions, normalized in such a way that the GCD is the total number of VSASMs
of size $5$. Each component, once divided by this GCD, counts VSASMs of size 5 with prescribed connectivity.
}

\subsec{Sum rule}
We are particularly interested here in the ``sum rule'' 
$Z_L(z_1,\ldots,z_L):=v\cdot\Psi_L(z_1,\ldots,z_L)
=\sum_\pi \Psi_{L,\pi}(z_1,\ldots,z_L)$
where we have set $q=\e{\pm 2i\pi/3}$ but retained the spectral parameter dependence.
$Z_L$ is a Laurent polynomial of degree width $2 L(L-1)/2$,
$2 (L-1)$ in each variable.
Applying $v$ on the left to Eqs.~\qkz{} (remembering that $v$ is left eigenvector of $\check R_i$, $K$
with eigenvalue $1$), we find immediately that $Z_L$ is a symmetric function of its arguments, and
that it is invariant by $z_i\to 1/z_i$. Next, we sum over $\pi$ in Eq.~\reca\ and obtain
\eqnn\recZ
$$\eqalignno{
Z_L(z_1,\ldots,z_L)_{z_L=q^2 z_{L-1}}=&Z_{L-2}(z_1,\ldots,z_{L-2})\times&\recZ\cr
&\times (1-q\,z_{L-1}\zeta^{-1})(\zeta-q^2z_{L-1}^{-1})
\prod_{i=1}^{L-2} (1-q\,z_{L-1}z_i^{-1})^2(z_i-q^2z_{L-1}^{-1})^2
}$$
where we have made use of Eq.~\exP.
The recurrence relation is started with $Z_0=Z_1=1$. $Z_L$, being a Laurent polynomial of degree width $2(L-1)$
in say $z_L$,
is entirely fixed by its value at $2L-1$ points. Eq.~\recZ\ (using the symmetries of $Z_L$)
provides us with the values of $Z_L$ at $z_L=q^{\pm 2} z_i^{\pm 1}$, which is more than enough.
Therefore we only need to find the unique solution of this recurrence.

Introduce the following determinant ratio: 
\eqn\defchi{
\chi_n(z_1,\ldots,z_n)={\det(z_i^{j+\lceil j/2\rceil-1}-z_i^{-j-\lceil j/2\rceil+1})_{1\le i,j\le n}
\over
\det(z_i^j-z_i^{-j})_{1\le i,j\le n}}
}
This function, a character of the symplectic group $Sp(2n)$,
already appears in \DF\ and in \Oka, as we shall discuss in more detail.
It is a symmetric Laurent polynomial of its arguments, invariant by $z_i\to 1/z_i$, 
of degree width $2\lceil n/2(n/2-1)\rceil$, 
and it satifies the recurrence relation:
$$\chi_n(z_1,\ldots,z_n)|_{z_n=q^2 z_{n-1}}=\chi_{n-2}(z_1,\ldots,z_{n-2})
\prod_{i=1}^{n-2}(1-q\,z_{n-1} z_i^{-1})(z_i-q^2 z_{n-1}^{-1})$$

We then easily find that
\eqn\Zdet{
Z_L(z_1,\ldots,z_L)=\chi_L(z_1,\ldots,z_L)\chi_{L+1}(z_1,\ldots,z_L,\zeta)
}
since it has the right degree and satisfies the recurrence.

\example{The first few sums:
$$\eqalignno{
Z_2&=z_1+z_2+\zeta+z_1^{-1}+z_2^{-1}+\zeta^{-1}\cr
Z_3&=(z_1+z_2+z_3+z_1^{-1}+z_2^{-1}+z_3^{-1})\times\cr
\times&\big(z_1z_2+z_1z_3+z_2z_3+z_1^{-1}z_2^{-1}+z_1^{-1}z_3^{-1}+z_2^{-1}z_3^{-1}
+(\zeta+\zeta^{-1})(z_1+z_2+z_3+z_1^{-1}+z_2^{-1}+z_3^{-1})\cr
&
+z_1z_2^{-1}+z_1z_3^{-1}+z_2z_3^{-1}+z_2z_1^{-1}+z_3z_1^{-1}+z_3z_2^{-1}+3\big)\cr
}$$}

Finally, we can take the homogeneous limit $z_i=1$. 
In all that follows we redefine the normalization of $Z$ so that
the smallest component is $\Psi_0=a^{\lfloor L/2\rfloor}$ as before: $Z_L(a)=3^{-L(L-1)/2} a^{\lfloor L/2\rfloor} Z_L(1,\ldots,1)$.
First we set $\zeta=1$ ($a=1$) as well.
The limit where parameters go to $1$ in Eq.~\defchi\ can be found, as always with these
types of determinants, by setting $z_i=\exp(\epsilon i)$ and letting $\epsilon$ go to zero.
This is strictly equivalent to using
Weyl's dimension formula for representations of $Sp(2n)$.
Define for future use $h_i=i+\lceil i/2\rceil-1$;
then $\chi_n(1,\ldots,1)=\prod_{1\le i<j\le n} {h_j-h_i\over j-i}\prod_{1\le i\le j\le n}{h_i+h_j\over i+j}$,
so that
$$\chi_n(1,\ldots,1) 3^{-\lceil n/2(n/2-1)\rceil} 
=\prod_{\scriptstyle 1\le k\le n-1\atop\scriptstyle 2|n-1-k} {\lfloor 3k/2+1\rfloor
(3k)!k!\over (2k+1)!(2k)!}=1,1,2,3,11,26,170,\ldots$$
where we have divided by the appropriate power of $3$ 
so that the resulting number has the following interpretation: it is,
for $n$ even, the number of Cyclically Symmetric Transpose Complement Plane Partitions
on a $n\times n\times n$ hexagon, and for $n$ odd, the number of Vertically Symmetric Alternating Sign
Matrices of size $n$. These numbers already appear as the sum of components in the case of closed boundary conditions
\refs{\PRdGN,\DF}.

Going back to $Z_L$, 
the powers of $3$ are cancelled by those coming from $\Psi_{L;0}(1,\ldots,1)=3^{L(L-1)/2}$,
and we conclude that
$$\sum_\pi \Psi_\pi = 
\prod_{k=1}^L {\lfloor 3k/2+1\rfloor
(3k)!k!\over (2k+1)!(2k)!}=1,2,6,33,286,4420,\ldots$$
which is the formula for the number of HVSASMs of size $2L+3$ conjectured by Mills \Rob\ and proved by Okada \Oka.

If we now reintroduce the $\zeta$ dependence, we find an expression which, as already stated, 
is conjecturally the weighted sum of VHSASMs where a weight $a$ is given to each pair of $-1$'s on 
the row below the horizontal symmetry axis to the left of the center.
In particular, in \dGR\ was formulated a conjecture about $\rho_L={1\over L}{\d\over\d a}\log Z_L|_{a=1}$
(which should be the average density of such $-1$'s).
Let us explain here how to compute this quantity.

One starts with the following identity related to $GL(N)$ characters, valid
for arbitrary integers $h_i$:
\eqn\schurid{
H_n(h_1,\ldots,h_N)=\sum_{i=1}^N {h_i^n\over \prod_{j(\ne i)} (h_j-h_i)}=
\cases{ 0& $n<N-1$\cr s_{n-N+1}(h_1,\ldots,h_N)&$n\ge N-1$}
}
where $s_n$ is the $GL(N)$ Schur function associated to the Young diagram with one single row of length $n$.
This is proved by considering the generating function $\sum H_n t^n$, using the fact that 
$H_0(h_1,\ldots,h_N,t^{-1})=0$ and concluding with the identity $\sum_{n\ge0} s_n(h_1,\ldots,h_N) t^n=
\exp\sum_{n\ge1}(\sum_i h_i^n)t^n/n$. Here we only need the case $n=N$, which, according
to Weyl's dimension formula, corresponds to computing
the first logarithmic derivative of a $GL(N)$ character wrt one parameter at the identity of the group,
the result being $\sum_i (h_i-i)$ which is nothing but the number of boxes of the Young diagram.

The corresponding ``symplectic'' identity can essentially be obtained by replacing $h_i$ with $h_i^2$.
In the present case we wish to compute the second derivative with respect to $\zeta$ of $\log \chi_{L+1}$
(since the first vanishes). We deduce, with $n=N=L+1$,
\eqn\sympschurid{
{\d^2\over\d\zeta^2}\log\chi_{L+1}(1,\ldots,1,\zeta)|_{\zeta=1}=
{1\over(L+1)(2L+3)} \sum_{i=1}^{L+1}(h_i^2-i^2)
}
where we have specialized to $h_i=i+\lceil i/2\rceil-1$.

Performing the summation: $\sum_{i=1}^{L+1}(h_i^2-i^2)=(L+1)\lfloor L/2\rfloor (5\lfloor (L+1)/2\rfloor+2)/3$,
and changing variables to $a=3/(\zeta+1+\zeta^{-1})$, not forgetting
the normalization $\Psi_0=a^{\lfloor L/2\rfloor}$, we obtain
\eqn\rhoid{
{\d\over\d a}\log Z_L|_{a=1}=\lfloor L/2\rfloor \left(1-{5\lfloor(L+1)/2\rfloor+2\over2(2L+3)}\right)
}
which is equivalent to the expression given in \dGR.


\def\ZUU{Z^{\rm UU}}
\newsec{Comments and conclusion}
\subsec{Relation to UUASM partition functions}
In view of the conjecture of \dG\ and its refinement in \dGR,
it is tempting to try to identify our sum rule with parameters
$Z_L(z_1,\ldots,z_L)$ with the partition function $\ZUU_L$
of Alternating Sign Matrices with two U-turn boundaries as introduced
in \Kup\ -- that is the partition function of the six-vertex model with its usual weights depending on
spectral parameters (and $q=\e{\pm 2i\pi/3}$) and certain special boundary conditions. The UUASMs are
a slightly more general class of ASMs than VHSASMs, but they are the right objects to consider once spectral
parameters are introduced; one can recover VHSASMs from UUASMs in the
homogeneous limit, as will be discussed below. 
The comparison reveals subtle differences. UUASMs being defined in even size only, we must discuss
the two parities separately:

\noindent $\star$ For even $L$, $\ZUU_L$ depends on $L$ spectral parameters,
just like ours (in the notations
of \Kup\ and \Oka, these are the square roots of our parameters $z_i$),
plus two boundary parameters, called $b$ and $c$. In \Oka, the case $b=c=q^{1/2}$ is
treated and the partition function is found to be equal (up to irrelevant prefactors) to 
$\chi_L(z_1,\ldots,z_L)\chi_{L+1}(z_1,\ldots,z_L,1)$. This coincides with $Z_L$ at $\zeta=1$.
It is in fact not too hard to check that if one sets $b^2=c^2=q/\zeta$, one also obtains the
dependence on $\zeta$. The amusing observation is that in the homogeneous limit $z_i=1$, the condition
$b=c=q^{1/2}$ selects VHSASMs (of size $2L+3$) among UUASMs; however setting $b^2=c^2=q/\zeta$ does not.
This provides an apparently non-trivial equality of the weighted sum
of UUASMs, where the weights are on the two U-turn boundaries (more precisely,
each U-turn gets a weight of $q-\zeta/q$ if it is the correct
orientation of the arrow for a VHSASM of size $2L+3$,
$1-\zeta$ if it is the wrong orientation)
and the weighted sum of VHSASMs where the weight $a=3/(\zeta+1+\zeta^{-1})$ is on the row below the
horizontal symmetry axis (for each pair of $-1$'s to the left of the center). As a check one can
consider the case $\zeta=q^2$, which selects among UUASMs the
VHSASMs one size less ($2L+1$): the conjectured equality implies that the number of VHSASMs
of size $2L+3$ with maximal power of $a$ is equal to the number of
VHSASMs of size $2L+1$, which can indeed be proven (in fact there is a
bijective proof, which is left as an exercise).

\noindent $\star$ For odd $L$, the situation is even more mysterious. $\ZUU_{L+1}$ depends
on $L+1$ parameters, that is one more than in our model. In \Oka, the case $b=c=q^{-1/2}$ (which,
in the homogeneous limit, selects VHSASMs of size $2(L+1)+1=2L+3$) is treated
and the result is that $\ZUU_{L+1}(z_1,\ldots,z_{L+1})\propto\chi_{L+1}(z_1,\ldots,z_{L+1})
\tilde\chi_{L+1}(z_1,\ldots,z_{L+1})$, where $\tilde\chi_{L+1}$ is a certain character of $O(2(L+1))$
which satisfies $\tilde\chi_{L+1}(z_1,\ldots,z_L,1)\propto \chi_L(z_1,\ldots,z_L)$.
Thus $\ZUU_{L+1}$ coincides with $Z_L(\zeta=1)$ on condition that one of its spectral parameter is set to $1$.
It is unclear how to reintroduce $\zeta$ in the partition function of UUASMs, and inversely
unclear how to incorporate the extra spectral parameter in our model.

\subsec{Limit $q\to 1$ and geometry}
The main focus of the present paper is on the special value $q=\e{\pm 2i\pi/3}$. However the rational limit of
the $q$KZ equation 
is also interesting. This corresponds to the limit $q\to 1$ (in \DFZJc\ the limit is indifferently
$q\to\pm1$, but this is because the shift $s=q^6$; here $s=q^3\to1$ imposes $q\to 1$), that is expanding
to first non-trivial order in $\hbar$ as $q=\e{-\hbar/2}$, $z_i=\e{-\hbar w_i}$.
In \DFZJc, it was found that solutions of the usual type $A$ (periodic boundary conditions)
rational $q$KZ equation based on Hecke algebra quotients
are related to the geometry (more precisely,
equivariant cohomology) of certain $sl(L)$ orbital varieties. This idea was generalized
in \DFZJd\ to Temperley--Lieb models with other boundary conditions which are apparently much closer to the present work;
the models were shown to be related to type $B$, $C$, $D$ orbital varieties. 
However a key difference is that the geometric interpretation
was ensured from the start by the fact that the $R$-matrix (both bulk and boundary) dependence
in the spectral parameter had a specific form which matched the Hotta construction of the Joseph
representation for orbital varieties. This resulted in slightly unusual variants of the Temperley--Lieb
algebra. Here the strategy is different: using the standard 1-boundary extended Temperley--Lieb algebra,
we are forced to use a certain boundary $R$-matrix whose geometric interpretation is not obvious.
Note that the example of \refs{\DFZJb,\KZJ} shows that more exotic varieties/geometric constructions 
than orbital varieties/Hotta construction can appear in integrable models.
We hope to come back to this question soon.

\subsec{Positivity conjectures}
In \KP\ it was observed that in the case of periodic boundary conditions, the components of solutions of the $q$KZ equation in which 
the variables $z_i$ are set to $1$ are polynomials with positive coefficients in the variable: (1) $\tau=-q-q^{-1}$ in the case
$s=q^6$; and (2) $\tau'=q^{1/2}+q^{-1/2}$ in the case $s=q^3$.
See also \DFb\ for an interesting conjecture on the combinatorial interpretation of the sum of homogeneous components for arbitrary $q$ in the case $s=q^6$.

We list here the first few homogeneous $\Psi_L(z_i=\zeta=1)$ in terms of $\tau'$ and after dividing by some common factors:
$$\eqalign{
\Psi_2&=(1,1)\cr
\Psi_3&=(1,2,2+\tau'^2)\cr
\Psi_4&=(1,3,2(3+\tau'^2),2+\tau'^2,5+3\tau'^2+\tau'^4,7+2\tau'^2)\cr
}$$
Clearly the positivity still seems to hold. Note that the special values considered above, namely $q=\e{\pm 2i\pi/3}$ and $q=1$, correspond
to $\tau'=1$ and $\tau'=2$ respectively.

\subsec{Other boundary conditions and prospects}
Several times in this article we have mentioned the similarities with the
work \DF. This is no surprise because the present model is a generalization
of the one considered in \DF: the Hamiltonian of the model with mixed
boundary conditions has an additional boundary term, and when the coefficient
$a$ in front of it is set to zero one recovers the Hamiltonian of the
model with closed boundary conditions. At the technical level, the main
difference is that the boundary term implies that one must include a
non-trivial boundary $R$-matrix satisfying the reflection equation. (Another difference,
irrelevant at the ``physical'' value $q=\e{\pm 2i\pi/3}$, is that the shift $s$ in the $q$KZ equation
turned out to be $q^3$ here, whereas it is $q^6$ in \DF).

Similarly, one could add another boundary term on the left, which
results in the model with open boundary conditions \MNdGB, an even broader
generalization. The analysis of the corresponding $q$KZ equation is more
involved, and is deferred to a future publication \CZJ.

\bigskip\goodbreak
\centerline{\bf Acknowledgments}
The author thanks L.~Cantini and J.-B.~Zuber for comments,
and P.~Di Francesco for pointing out Ref.~\dGR; and acknowledges the support
of European Marie Curie Research Training Networks ``ENIGMA'' MRT-CT-2004-5652, ``ENRAGE'' MRTN-CT-2004-005616,
and of ANR program ``GIMP'' ANR-05-BLAN-0029-01.

\listrefs
\end